  \documentstyle[twocolumn]{article}
\makeatletter
\newdimen\normalarrayskip              
\newdimen\minarrayskip                 
\normalarrayskip\baselineskip \minarrayskip\jot
\newif\ifold             \oldtrue            
\def\arraymode{\ifold\relax\else\displaystyle\fi} 
\def\eqnumphantom{\phantom{(\theequation)}}     
\def\@arrayskip{\ifold\baselineskip\z@\lineskip\z@
     \else
     \baselineskip\minarrayskip\lineskip2\minarrayskip\fi}
\def\@arrayclassz{\ifcase \@lastchclass \@acolampacol \or
\@ampacol \or \or \or \@addamp \or
   \@acolampacol \or \@firstampfalse \@acol \fi
\edef\@preamble{\@preamble
  \ifcase \@chnum
     \hfil$\relax\arraymode\@sharp$\hfil
     \or $\relax\arraymode\@sharp$\hfil
     \or \hfil$\relax\arraymode\@sharp$\fi}}
\def\@array[#1]#2{\setbox\@arstrutbox=\hbox{\vrule
     height\arraystretch \ht\strutbox
     depth\arraystretch \dp\strutbox
     width\z@}\@mkpream{#2}\edef\@preamble{\halign \noexpand\@halignto
\bgroup \tabskip\z@ \@arstrut \@preamble \tabskip\z@ \cr}%
\let\@startpbox\@@startpbox \let\@endpbox\@@endpbox
  \if #1t\vtop \else \if#1b\vbox \else \vcenter \fi\fi
  \bgroup \let\par\relax
  \let\@sharp##\let\protect\relax
  \@arrayskip\@preamble}
\def\eqnarray{\stepcounter{equation}%
          \let\@currentlabel=\theequation
          \global\@eqnswtrue
          \global\@eqcnt\z@
          \tabskip\@centering
          \let\\=\@eqncr
          $$%
 \halign to \displaywidth\bgroup
    \eqnumphantom\@eqnsel\hskip\@centering
    $\displaystyle \tabskip\z@ {##}$%
    &\global\@eqcnt\@ne \hskip 2\arraycolsep
     $\displaystyle\arraymode{##}$\hfil
    &\global\@eqcnt\tw@ \hskip 2\arraycolsep
     $\displaystyle\tabskip\z@{##}$\hfil
     \tabskip\@centering
    &{##}\tabskip\z@\cr}
\makeatother

\begin{document}
\newcommand{\beq}{\begin{equation}}
\newcommand{\eeq}{\end{equation}}

\begin{titlepage}
\begin{center}
\begin{flushright}
{\small November 19, 1999}
\end{flushright}

 \vspace{.5cm}

{\Large {\bf The world is not enough}} \vspace{.5cm}

{\bf Luis A. Anchordoqui$^a$ and Santiago E. Perez
Bergliaffa$^b$}\\ \vspace{.5cm} {\it $^a$Department of Physics,
Northeastern University, Boston, Massachusetts 02115}\\ {\it
$^b$Centro Brasileiro de Pesquisas Fisicas, Rua Xavier Sigaud,
150, CEP 22290-180, Rio de Janeiro, Brazil}

\end{center}

\vspace{1.5cm}

\begin{center}
{\bf Abstract}
\end{center}

We show that the 5-dimensional model introduced by Randall and
Sundrum is (half of) a wormhole, and that this is a general result in models
of the RS type. We also  discuss the gravitational trapping of a scalar 
particle in 5-d spacetimes. Finally, we present a simple model of brane-world
cosmology in which the background is a static anti-de Sitter
manifold, and the location of the two 3-branes  is determined by
the technique of ``surgical grafting''.

\hfill


\vspace{2cm}

\begin{center}

{\it To be published in Physical Review D.}

\end{center}

\end{titlepage}

In recent years, high energy particle physics and gravity have
been reconciled by means of the so--called brane worlds.
\cite{review}. From the phenomenological perspective these worlds
provide an economic explanation of the hierarchy between the
gravitational and electroweak mass scales. In particular, it has
been suggested that the fundamental Plank scale $M_*$ can be
lowered all the way to TeV scale \cite{susy} by introducing extra
dimensions \cite{a-h}. In this framework the Standard Model (SM)
is confined to a 3-brane \cite{M} whereas gravity propagates
freely through the extra dimensions. In such higher--dimensional
models spacetime is usually taken to be the product of a
4--dimensional spacetime and a compact $n$-manifold. Consequently,
the observed Plank scale is related to the higher dimensional
scale gravity $M_*$ through the relation $M_{\rm pl}^2 =
M_*^{n+2}V_n$, being $V_{n}$ the volume of the compact dimensional
space. Unfortunately, the presence of large extra dimensions does
not necessarily provide a satisfactory resolution of the hierarchy
problem, which reappears in the large ratio between $M_*$ and the
compactification scale $\mu_c = V_n^{-1/n}$. Following the idea
that SM fields may live in a 3-brane, Randall and Sundrum (RS)
presented an alternative solution with the shape of a
gravitational condenser: two branes of opposite tension (which
gravitationally repel each other) stabilized by a slab of anti-de
Sitter (AdS) space \cite{rs1}. In this model the extra dimension
is strongly curved, and the distance scales on the brane with
negative tension are exponentially smaller than those on the
positive tension brane.

Two different scenarios might be analyzed with this solution. We
could assume that our world is the brane with negative tension, so
as to obtain a correct ratio between the Plank and weak scales
without the need to introduce a large hierarchy between $M_*$ and
$\mu_c$ \cite{rs1}. Alternatively, if the visible brane is the one
with positive tension, the configuration does not solve the
hierarchy problem. However, the second brane can be moved to
infinity and Newton's law is correctly reproduced on the
brane--world in spite of the non-compact extra dimension
\cite{rs2}. In both cases, Poincar\'e invariance on the 3-brane
was assumed \cite{blackhole}. Much of the recent work in this area
has been devoted to lift this restriction, in order to see under
what conditions it is possible to obtain the standard
Friedmann-Robertson-Walker cosmological scenario on the visible
brane \cite{cosmology}. In this Brief Report we shall show that
the spacetime obtained in \cite{rs1} has the structure of a
wormhole. We shall also show here that any 5-dimensional
non--factorizable geometry (i.e., one in which the components of
the metric tensor depend on the coordinate of the compactified
extra dimension) is accompanied by unavoidable violations of the
null energy condition. Furthermore, we shall see that this is a
direct consequence of the fact that within these models the
visible brane is located at the throat of the wormhole. After
studying these relations between wormholes and the phenomenology
of RS spacetimes,  we 
discuss the mechanism of gravitational trapping for a scalar particle 
in a general static spacetime with diagonal 3-metric. Afterwards
we work out a new solution of Einstein's
equation given by two slices of AdS spacetimes glued together at
the location of the branes (one of which is at the wormhole's
throat). Some features of the cosmology of this systems are
presented. We close with a discussion. \\[0.6cm]

{\em Do we live in a wormhole throat?} Morris and Thorne showed
\cite{motho} that the basic feature of a wormhole is that the
``flaring-out condition'' must be satisfied at its throat. A
throat is a closed spatial hypersurface such that one of the two
future-directed null geodesic congruences orthogonal to it is just
beginning to diverge on or near the surface. Stated
mathematically, the expansion $\theta_\pm$ of one of the two
orthogonal null congruences vanishes on the surface: $\theta_+ =
0$ and/or $\theta_- = 0$, and the rate-of-change of the expansion
along the same null direction $(u_\pm)$ is positive-semi-definite
at the surface: $d\theta_\pm/du_\pm \geq 0$ \cite{h-v2}. Before
showing that the solution given by RS \cite{rs1} can be
interpreted as part of a 5--dimensional wormhole with its throat located
on the negative tension brane, we recall the set up of the RS
model.

Let us start from the 5-dimensional action,
\begin{equation}
S = S_{\rm gravity} + S_{vis} + S_{hid},
\end {equation}
where
\begin{equation}
S_{\rm gravity} = \int d^4x \int_0^{\pi r_c} dy \sqrt{G}\;
\{-\Lambda + 2 M_*^3 R\},
\end{equation}
\begin{equation}
S_{\rm vis} = \int d^4x \sqrt{-g_{\rm vis}} \;\{{\cal L}_{\rm vis}
- V_{\rm vis} \}, \label{avis}
\end{equation}
and
\begin{equation}
S_{\rm hid} = \int d^4x \sqrt{-g_{\rm hid}}\; \{{\cal L}_{\rm hid}
- V_{\rm hid} \}. \label{ahid}
\end{equation}
Here $R$ stands for the 5-dimensional Ricci scalar in terms of the
metric $G_{AB}$, and $\Lambda$ is the 5-dimensional cosmological
constant. The coordinate $y$ is the extra dimension, and its range
is $[0,\pi r_c]$, where $r_c$  is the compactification radius. RS
work on the space $S^1/{\bf Z_2}$. Eqs.(\ref{avis}) and
(\ref{ahid}) represent the action of 3-branes located at the
orbifold fixed points at $y = 0, \pi r_c$. From the Lagrangians in
(\ref{avis}) and (\ref{ahid}) it has been separated out a constant
vacuum energy which acts as a gravitational source even in the
absence of particle excitations. In what follows we shall drop the
${\cal L}$ terms since they are not relevant in determining the
classical 5-dimensional metric in the ground state. The Einstein
equation for the above action reads,\footnote{Capital Latin
indices run from 0 to 4 and refer to the entire 5-dimensional
spacetime, Greek indices run from 0 to 3 and refer to the brane
sub-spacetime; Latin indices from the middle of the alphabet
$(i,j,k,\dots)$ run from 1 to 4 and refer to constant $t$ slices;
Latin indices from the beginning of the alphabet $(a,b,c,\dots)$
will run from 1 to 3 and will be used to refer to the 3-brane. As
usual, $\eta_{\mu\nu}$ stands for the four dimensional Minkowski
spacetime.}

\begin{eqnarray}
\sqrt{-G} \,(R_{AB} - \frac{1}{2}\, G_{AB}\, R) & = &  -
\frac{1}{4M_*^3}\,\, [\,\Lambda \,\sqrt{-G}\, G_{AB} \nonumber \\
 & + & V_{\rm vis} \,\, \sqrt{-g_{\rm vis}} \,\,g_{\mu\nu}^{\rm
vis}\,\, \delta_A^\mu \,\delta_B^\nu \, \delta(y - y_0) \nonumber
\\  & + & V_{\rm hid} \,\, \sqrt{-g_{\rm hid}} \,\,g_{\mu\nu}^{\rm
hid}\,\, \delta_A^\mu \,\delta_B^\nu \,\delta(y)].
\label{einstein}
\end{eqnarray}
The line element
\begin{equation}
ds^2 = e^{-2 k |y|} \eta_{\mu\nu} dx^\mu dx^\nu + dy^2,
\label{lisa-metric}
\end{equation}
will be a solution of Eq.(\ref{einstein}) provided that the
boundary and bulk cosmological terms are related by a single scale
$k$,
\begin{equation}
V_{\rm hid} = - V_{\rm vis} = 24\, M_*^3\,k, \,\,\,\,\,\, \Lambda=
-24\,M_*^3\,k^2.
\end{equation}

We turn now to the analysis of the null geodesic congruences. We
recall that in any static 5-dimensional spacetime every focusing
(defocusing) of null geodesic congruences starts at a closed
3-dimensional hypersurface of maximal (minimal) area that, without
loss of generality, can be located within a single constant-time
spatial slice \cite{h-v}. To describe the behavior of the
brane-worlds, we conveniently adopt a Gaussian normal coordinate
system in the neighborhood of each brane. We shall denote the
3-dimensional hypersurface swept out by each brane by $\Sigma$.
Let us introduce a coordinate system $x^a_\perp$ on $\Sigma$. Next
we consider all the geodesics which are orthogonal to $\Sigma$,
and choose a neighborhood $N$ around $\Sigma$ so that any point $p
\in N$ lies on one, and only one, geodesic. The first three
coordinates of $p$ are  determined by the intersection of this
geodesic with $\Sigma$. The full set of spatial coordinates is
then given by $x^i=(x^a_\perp;\ell)$, wherein the hypersurfaces
$\Sigma$ under consideration are taken to be at $\ell=0$, so that
the spacetime metric reads
\begin{equation}
- e^\phi dt^2 + ^{(4)}\!g_{ij} dx^i dx^j = \, -e^\phi dt^2+
^{(3)}\!g_{ab} dx^a dx^b + d\ell^2,
\end{equation}
with $\phi$ the redshift function. The extrinsic curvature of each
3-surface is defined by
\begin{equation}
K_{ab} = \frac{1}{2} \, \frac{\partial g_{ab}}{\partial \ell}.
\end{equation}
If we compute the variation in the area of $\Sigma$ obtained by
pushing the surface at $\ell=0$ out to $\ell = \delta \ell(x)$ we
get
\begin{equation}
\delta A (\Sigma) = \int \sqrt{^{(3)}\! g} \;{\rm tr} (K)\; \delta
\ell(x) \; d^3x.
\end{equation}
Since this expression must vanish for arbitrary $\delta \ell(x)$,
the condition for the area to be extremal is simply tr$(K)=0$. In
the case of the RS model described by Eq.(\ref{lisa-metric}) the
extrinsic curvature is given by
\begin{equation}
K_{ab} = -k \,\,{\rm sg} (y)\,\, e^{-2k|y|} \,\eta_{ab}
\label{excur}
\end{equation}
Straightforward calculations show that the 3-sections located at
$y=0$ and $y= \pm \pi r_c$ are extremal. For the area to be
minimal the additional constraint $\delta^2 A(\Sigma) \geq 0$ is
required. Equivalently,
\begin{equation}
\frac{\partial \,{\rm tr} (K)}{\partial \ell} \geq 0.
\end{equation}
For the RS spacetime we get
\begin{equation}
\frac{\partial {\rm tr}(K)}{\partial y} = - 3 \,k \,\delta(y),
\end{equation}
then it is easily seen that the brane located at $y=0$ represents
a hypersurface of maximal area. The 3-surface at $y=\pm \pi r_c$
is instead a hypersurface of minimal area, {\em i.e.} a wormhole
throat. Note that, due to the double orbifolding, the two different sides of the throat are actually topologically identified. Consequently, the RS model can be thought of as ``half a wormhole''.

It can be shown that any extension of the standard 4-dimensional
world using a compactified non--factorizable dimension will posses
hypersurfaces of maximal and minimal area. This assertion follows
from the fact that the area of the 3-branes depends only on the
extra coordinate. Then, as a function of the extra coordinate it
must have an absolute maximum and an absolute minimum in the
finite closed interval, and so forth must violate the null energy
condition. Clearly, this is not the case with an infinite extra
dimension \cite{lr}.

Putting  all this together, the huge hierarchy between the
gravitational and weak scales seems to be the result of the change
of the scales throughout the null geodesic congruence of a
compactified dimension. \\[0.6cm]

{\it Gravitational trapping}. As we mentioned above, in the RS
framework nongravitational fields must be completely confined to
the brane by some mechanism. On the other hand, the zero mode of
gravity is localized on the brane and higher modes propagate
freely in the extra dimension. Several confining mechanisms have
been proposed in the past. For instance, in \cite{rubakov} it was
shown that a potential well (originating in the nonlinearities of
the eqs. of motion of a scalar field) can force scalar and spin
1/2 particles to live on the brane. Later, it was shown in
\cite{dvali} how massless gauge fields can be localized on a brane
due to the dynamics of the gauge field outside the wall. Here we
explore the possibility that particles can be trapped on the
3-brane by the sole action of gravity. This idea has been analyzed
before in \cite{kk-visser,gib,ruso1,ruso2}. We will partially
follow the work by Visser \cite{kk-visser}. To see how the
gravitational trapping works in our geometric approach, we shall
briefly discuss the localization of a massless scalar field in
5-dimensional static spacetimes. The equation of motion is given
by
\begin{equation}
\partial_A (\sqrt{G} \,G^{AB} \partial_B \Psi) = 0. \label{KG1}
\end{equation}
In order to solve Eq.(\ref{KG1}), we assume that the redshift
function depends only the extra coordinate, and that the metric on
the brane-world is diagonal. This assumption implies that ${\rm
tr} (K) \equiv \partial_\ell g^{^{(3)}}/2\,g^{^{(3)}}$, where
$^{(3)}g$ is the determinant of the 3-metric. Moreover, we adopt
for $\Psi$ the following {\em Ansatz}\/:
\begin{equation}
\Psi = e^{-i p_\mu x^\mu} \,\psi(\ell),
\end{equation}
(with $p_\mu p^\mu = - m^2$) which represents a travelling wave in
the four dimensional submanifold. Using the change of variables
$\psi(\ell) = u(\ell) \exp \left(- 1 / 2 \int^\ell \{\phi'(z) +
{\rm tr} [K(z)]\}\, dz \right )$, one gets an equation for
$u(\ell)$
\begin{equation}
u''(\ell) + \left\{e^{-2\phi} m^2 - \frac 1 4 [\phi ^{'} + {\rm
tr} (K) ]^2 -\frac 1 2 \phi '' - \frac 1 2 [{\rm tr}(K)]'\right\}
u(\ell) = 0. \label{equ}
\end{equation}
By means of the transformations $z = z(\ell)$ and $u(z) = \gamma
(z) \sigma(z)$, this equation can be re-written as a
one-dimensional Schr\"odinger-like equation \beq
\frac{d^2\sigma}{dz^2} + V(z) \,\,\sigma = -m^2 \sigma,
\label{sch} \eeq with \beq V(z) = \frac 1 \gamma
\frac{d^2\gamma}{dz^2} + \frac{d(\ln\gamma)}{dz}
\frac{z''}{(z')^2}  - \frac{[\phi ^{'} + {\rm tr}(K)]^2 + 2 \phi
'' + 2 [{\rm tr}(K)]' } {4\,(z')^2}, \eeq and $ z(l) = \int
e^{-\phi} dl $, $\ln \gamma =  1/ 2 \int \phi' e^\phi z' dl$ (the
prime denotes derivative w.r.t $\ell$). The potential $V(z)$ will
determine whether the scalar particle is localized on the brane,
and will also yield the 4-dimensional mass spectrum. Let us remark
that this equation is valid for any static model in which the
3-metric is diagonal (e.g. the ``Visser gauge'' can be obtained by
fixing $K_{ab} = 0$ \cite{kk-visser}). In the specific case of the
RS model we have \beq z(\ell ) = \frac 1 k {\rm sg}(\ell )
(e^{k|\ell |} - 1), ~~~~~~~~~~~~~~~~\gamma(z) = (1+k|z|)^{-1/2},
\nonumber \eeq and \beq V(z) = \frac{3}{2} k \, \delta (z) -
\frac{15}{4} \frac{k^2}{(1+k|z|)^2}. \eeq The zero-mass solution
takes the form $\sigma_0(z)= k^{-1}(1+ k |z|)^{-3/2}$. As in the
case of gravitons Eq. (\ref{sch}) gives rise to a tower of
continuum states with $m^2>0$ \cite{rs2}. A general discussion on
the mass spectrum can be found in \cite{bajc}. The analysis
carried out here in terms of the extrinsic curvature and the
redshift function can be straightforwardly generalized to nonzero spin fields. The case of particles with spin 1, 1/2 and 3/2 in
the RS model was studied in \cite{bajc}.\\[0.6cm]

{\em The cosmological solution.} We shall now construct a solution
of Eq.(5) by connecting ``incomplete'' AdS spaces according to the
technique introduced by Visser \cite{visser-book}, we shall take
two copies of 5-dimensional AdS spacetimes and join them with the
junction condition formalism \cite{israel}. Namely, first remove
from each AdS spacetime identical regions of the form ${\cal R}
\times \Omega$, where $\Omega$ is a 4-dimensional compact
spacelike hypersurface and ${\cal R}$ is a timelike straight line.
Identify now these two incomplete spacetimes along the timelike
boundaries ${\cal R} \times \partial \Omega$. The resulting
spacetime ${\cal M}$ is geodesically complete and posses two AdS
asymptotic regions connected by a wormhole. The throat of the
wormhole is just the junction $\partial \Omega$ at which the two
original AdS spacetimes are identified.

For definiteness, the coordinates on $\partial \Omega$ can be
taken to be the angular variables ($\chi$, $\theta$, $\phi$) which
are always well defined, up to an overall rotation, for a
spherically symmetric configuration. We shall refer again to a
Gaussian normal coordinate system in the neighborhood of the
throat,
\begin{equation}
ds^2 = - (1-\lambda r^2)\, dt^2 + dy^2 + r^2\, d\Omega_3^2,
\end{equation}
where
\begin{equation}
\frac{dr}{dy}= \sqrt{1 - \lambda  r^2}, \label{gauss}
\end{equation}
$\lambda = \Lambda/24M_*^3$, $r$ and $t$ denote the values of the
corresponding AdS coordinates, and $d\Omega_3^2$ stands for the
line element of the three sphere.

Associating positive values of $G_{AB}$ to one side of the throat
and negative values to the other side, without loss of generality
the metric in the neighborhood of the brane can be written as,
\begin{equation}
G_{AB}  = \Theta (y - y_0) \,\, G_{AB}^+  +\,\, \Theta (-y + y_0)
\,\,G^-_{AB}.
\end{equation}
At the boundary of the two regions, the energy momentum tensor can
be expressed in terms of the jump of the extrinsic curvature
${\cal K}^A_B = K^{A\,+}_B - K^{A\,-}_B$,
\begin{equation}
\left. K_{AB}^\pm = \frac{1}{2} \frac{\partial G_{AB}}{\partial y}
\right|_{y = y_0^\pm} = \frac{1}{2} \frac{\partial
G_{AB}^\pm}{\partial y}.
\end{equation}

At this stage, it is convenient to introduce  the coordinate which
denotes the proper time as measured along the surface of the
wormhole throat $\tau$. Then, for the dynamic case, the geometry
is uniquely specified by a single degree of freedom, the radius of
the throat $a(\tau)$. Across the boundary, the jumps of the
extrinsic curvature are given by \cite{kim} (dots indicate
derivative with respect to $\tau$)
\begin{equation}
{\cal K}^\tau_\tau = \frac{2\,(\ddot{a} - \lambda
a)}{\sqrt{1-\lambda a^2 + \dot{a}^2}},
\end{equation}
and
\begin{equation}
{\cal K}^\chi_\chi = {\cal K}^\theta_\theta = {\cal K}^\phi_\phi =
\frac{2}{a} \,\sqrt{1-\lambda  a^2 + \dot{a}^2}.
\end{equation}
Replacing in Eq.(\ref{einstein}) we obtain,
\begin{equation}
 V_{\rm vis}  \,g_{\mu\nu}^{\rm vis}\,\, \delta_A^\mu \,\delta_B^\nu \,  =
 4 M_*^3\, [{\cal
 K}_{AB}\, - {\rm tr} ({\cal K}) \,g_{\mu\nu}^{\rm vis}\,\, \delta_A^\mu
\,\delta_B^\nu \, ],
\end{equation}
or equivalently,
\begin{equation}
V_{\rm vis} = - 24 M_*^3 \,\,\frac{\sqrt{1 - \lambda a^2 +
\dot{a}^2}}{a},
\end{equation}
\begin{equation}
V_{\rm vis} = -8M_*^3 \,\,\frac{2 - 3\lambda a^2 + 2 \dot{a}^2 +
\ddot{a} a}{a \,\sqrt{1- \lambda a^2 + \dot{a}^2}}.
\label{tension}
\end{equation}

Now, Eq.(\ref{einstein}) together with the covariant conservation
of the stress energy determine the classical dynamics of the
system \cite{quantum},
\begin{equation}
\dot{a}^2 - \left[ \frac{\Lambda}{24M_*^3} + \left( \frac{V_{\rm
vis}}{24\,M_*^3}\right) ^2 \right]\, a^2 = -1. \label{009}
\end{equation}
After integration we get,
\begin{equation}
a = a_0 \,\, {\rm cosh} \{ a_0^{-1}\, \tau\},
\end{equation}
where $a_0 = (\lambda+(V_{\rm vis}/24M_*^3)^2)^{-1/2}$ is the
minimum radius of the brane. Contrary to what happens in the RS
model, if $\lambda <0$ and $|\lambda| \geq (V_{\rm
vis}^2/24M_*^3)$, $a$ has no real solution. However, if either
$\lambda >0$, or $\lambda<0$ and $|\lambda| < (V_{\rm
vis}^2/24M_*^3)$, the wormhole does not  collapse but it is
bounded by a minimum size $a_0$ \cite{malda}. The expression for
the Hubble constant in this solution is
\begin{equation}
H^2 = -\frac{1}{a^2} + \frac{\Lambda}{24M_*^3} +
\left(\frac{V_{\rm vis}}{24 M_*^3}\right)^2. \label{hubble}
\end{equation}
Its behavior is similar to the one obtained for the RS models,
i.e., $H \propto V_{\rm vis}$ rather than $V_{\rm vis}^{1/2}$ as
in conventional cosmology. However, as the world approaches to the
minimum size the expansion tends to zero. Furthermore, unlike the
standard spherical Robertson-Walker case this world experience an
everlasting expansion.

To find out the behavior of the surface of maximal area one must
redefine the jump in the extrinsic curvature ${\cal K}^A_B =
K^{A\,-}_B - K^{A\,+}_B$, and afterwards repeat {\em mutatis
mutandis} the entire computation. It is easily seen that except
for the sign of the vacuum energy which results flipped $(V_{\rm
hid} >0)$, one can just replace the sub-index vis by hid in every
expression. The expansion rate of the hidden brane is then given
by,
\begin{equation}
 \left( \frac{\dot{b}}{b} \right)^2= - \frac{1}{b^2} + \frac{\Lambda}{24M_*^3} +
 \left(\frac{V_{\rm hid}}{24 M_*^3} \right)^2, \label{hubbleb}
\end{equation}
where $b$ is the radius of the hidden brane.\footnote {Let us
mention that the dynamics of a domain wall in the RS-$AdS_5$ bulk
was analyzed in (Kraus Ref. \cite{cosmology}). With the help of
the same formalism the author found a rather similar solution
where the bulk cosmological constant is not exactly cancelled by
the one on the brane.}\\[0.6cm]

{\em Discussion}. Let us suggest a possible way to understand the
hierarchy problem in the scenario presented here. To preserve the
stability of the system, we need to impose a constraint on the
relation between $\delta a$ and $\delta b$, the increments in the
visible and hidden radii during an interval of time as measured by
synchronized clocks on each brane. If this constraint is given,
the strategy for solving the hierarchy problem is as follows.
First one has to select $\delta a$ and $\delta b$ so as to obtain
the correct relation between the mass scales, then
Eq.(\ref{gauss}) will fix $\lambda$ to obtain that hierarchy. The
value of the present Hubble parameter together with Eq.
(\ref{hubble}) set $V_{\rm vis}$. Finally, from Eq.
(\ref{hubbleb}) one can determine the value of $V_{\rm hid}$.

Our solution can be regarded as a 5-dimensional spacetime (with
two brane worlds located at fixed coordinates in the internal
dimension) propagating through a static AdS background manifold. A
more general case would be given by lifting the static restriction
on the background. This may have some important consequences.
Consider for instance an expanding 3-dimensional spacelike
hypersurface embedded in a 4+1-dimensional spacetime. At each
point on the hypersurface there are two null vectors orthogonal to
the hypersurface, and associated to these two null vectors there
exist two null geodesic congruences that are well defined on an
open neighborhood of the hypersurface. While the surface
propagates along the null geodesic congruence of the extra
dimension the mass scale changes. Hence, if the speed of the
surface (along the internal dimension) is comparable to its
expansion rate, the signals that travel into the world may display
an additional redshift or blueshift. In the former case the
observers on the brane would interpret its motion as an
inflationary scenario.

We have proved above that the geometrical structure of the
manifold given by RS 
corresponds to half of a wormhole geometry, and its mirror image. 
We also showed that 
any geometry with a compactified non-factorizable dimension will 
inevitably posses hypersurfaces of maximal and minimal area.
Now, the
dependence of the extrinsic curvature of the RS solution with the
extra coordinate, given by  Eq.(\ref{excur}), coincides with that
of the warp factor in the RS metric. In turn, this factor is
responsible for the renormalization of the mass that solves the
hierarchy problem in the RS model. This coincidence may be
suggesting a relation between the extrinsic curvature and the
change in the mass scale. We will pursue this line of research
somewhere else.

\section*{Acknowledgments}We have benefited from discussions with
Carlos Nu\~nez, Per Kraus, Yogi Srivastava, and John Swain. This
work was partially supported by CONICET.


\begin{thebibliography}{99}
\bibitem{quantum} For a recipe to analyse the quantum stability in
the minisuperspace approximation the reader is referred to Part V
of Ref. \cite{visser-book}.
\bibitem{malda} Notice that although the  solution with $\lambda >0$ (de Sitter
space) is  well behaved, we have always taken the negative
cosmological term in the bulk. Our motivation for this sign choice
is that such models can be naturally obtained by compactifying the
II B string theory on AdS$_5$ $\times S^5$. See J. Maldacena, Adv.
Theor. Math. Phys. {\bf 2}, 231 (1998).
\end{thebibliography}
\end{document}